\documentclass[%
 reprint,
 amsmath,amssymb,
 aps,
]{revtex4-2}

\usepackage{graphicx}
\usepackage{dcolumn}
\usepackage{bm}

\usepackage{xcolor, soul} 

\begin{document}

\preprint{APS/123-QED}

\title{Emergent magnetic monopole and dipole screening by proximity effect with noble metal}

\author{Fernando F. Martins}
 \affiliation{Laboratory of Spintronics and Nanomagnetism ($LabSpiN$), Departamento de F\'{i}sica, Universidade Federal de Vi\c{c}osa, 36570-000 - Vi\c{c}osa - Minas Gerais, Brazil.}
 \author{Te\^{o}nis S. Paiva}
 \affiliation{Laboratory of Spintronics and Nanomagnetism ($LabSpiN$), Departamento de F\'{i}sica, Universidade Federal de Vi\c{c}osa, 36570-000 - Vi\c{c}osa - Minas Gerais, Brazil.}
\author{Daniel G. Duarte}%
\affiliation{Laboratory of Spintronics and Nanomagnetism ($LabSpiN$), Departamento de F\'{i}sica, Universidade Federal de Vi\c{c}osa, 36570-000 - Vi\c{c}osa - Minas Gerais, Brazil.}%
\author{Jo\~{a}o H. Rodrigues}%
\affiliation{Núcleo de F\'{i}sica, Instituto Federal de Minas Gerais - Campus Bambu\'{i}, 38900-000 Bambu\'{i}, Minas Gerais, Brazil}%
\author{Lucas A. S. M\'{o}l}%
\affiliation{Laborat\'{o}rio de Simula\c{c}\~{a}o, Departamento de F\'{i}sica, ICEx, Universidade Federal de Minas Gerais, 31720-901 Belo Horizonte, Minas Gerais, Brazil}%
\author{Jerome Borme}%
\affiliation{INL-International Iberian Nanotechnology Laboratory,
4715-330, Braga, Portugal}%
\author{Paulo P. Freitas}%
\affiliation{INL-International Iberian Nanotechnology Laboratory,
4715-330, Braga, Portugal}%
\author{Clodoaldo I. L. de Araujo}%
\email{dearaujo@ufv.br}
\affiliation{Laboratory of Spintronics and Nanomagnetism ($LabSpiN$), Departamento de F\'{i}sica, Universidade Federal de Vi\c{c}osa, 36570-000 - Vi\c{c}osa - Minas Gerais, Brazil.}%

\date{\today}

\begin{abstract}
In this letter we present emergent screening of magnetic monopole and dipole by the presence of 20nm aluminum cover layer. Our results were obtained in base of magnetic atomic force measurements, performed after external magnetic field steps application. We show that the evolution of magnetization and monopole population is affected by the aluminum presence and attribute that phenomena to the proximity effect, which is responsible for the magnetization vanish of the first atomic layers at the interface. Using experimental values to estimate the decrease in the nanomagnetic dipole value used in an emergent excitation model and in the switching field distribution heterogeneity used in simulations, we observe a very good agreement among experimental and simulation results. The presented emergent screening could be used in new $ASI$ geometries for thermodynamic activation or proposition of devices with selective magnetic monopole mobility.    

\end{abstract}

\maketitle


The quantization of electronic charge could just be explained so far by the existence of magnetic monopoles\cite{dirac1931quantised}. Although, its scarcity to be probabilistic detected \cite{polchinski2004monopoles} and difficulty of creation in particle accelerators due to its predicted high mass\cite{aad2020search}, still leaves the existence of such elementary particles unproven. In condensed-matter systems, magnetic monopoles were first reported as low energy emergent quasiparticles in ferromagnetic crystals \cite{fang2003anomalous} and have been mostly observed in pyrochlore crystals, denominated natural spin ices by resembling geometric frustration and residual entropy of water ice\cite{castelnovo2008magnetic}. In those crystals, at very low temperatures in the range of 0.6-2K, magnetic field was used to break symmetry and align the non energetic strings connecting monopoles, allowing its transport to generate magnetricity \cite{bramwell2009measurement}. The original proposition \cite{mol2009magnetic} and experimental observation at room temperature \cite{morgan2011thermal} of magnetic monopole quasiparticles in square array of nanomagnets, inspired a plethora of alternative designs proposition presenting singular properties\cite{qi2008direct,mol2012extending,ribeiro2017realization,loreto2015emergence,nisoli2017deliberate,gilbert2014emergent,macedo2018apparent}. 
Despite the fact that easy sample nanofabrication brings very interesting implications for novel technological applications, traditional artificial spin ice ($ASI$) systems present two main obstacles to achieve that purpose. One is the high Curie temperature of conventional Permalloy nanomagnets \cite{kapaklis2012melting}, that prevents large scale ground state achievement by thermal effects\cite{silva2012thermodynamics,wysin2015order} and the other is the energetic string connecting Nambu monopoles\cite{silva2013nambu,morley2019thermally} in bidimensional $ASI$, making them not free for magnetricity as in its natural counterpart. Those problems were circumvented so far by utilization of thinner nanomagnets, close to superparamagnetic regime, lowering Curie temperatures\cite{anghinolfi2015thermodynamic,farhan2016thermodynamics},  or utilization of low dimension in size arrays to achieve full ground state\cite{zhang2019understanding}. Both strategies could be fundamental to record and erase excitations in future technological applications. In order to surmount the strong connection between opposite monopoles in two dimensional $ASI$, systems close to degeneracy with vanishing string energy by lattice stretch have been proposed \cite{nascimento2012confinement,loreto2018experimental,gonccalves2019tuning}. More recently, the theoretical prediction of monopole freedom \cite{mol2010conditions} in real degenerate systems of three dimensional $ASI$ \cite{moller2006artificial} was experimentally realized \cite{perrin2016extensive} and a magnetic monopole plasma could be characterized\cite{farhan2019emergent}. All those advances together with possibilities of information record by specific nanomagnet manipulation, with support of external field applied just below the nanomagnetic coercivity\cite{nisoli2018write}, opened a pave for technological applications such as logic gates\cite{arava2019engineering} or memristors\cite{caravelli2019artificial} for neuromorphic applications. 
In this letter we aim to introduce an alternative way to manipulate monopole density and mobility in square $ASI$. Such manipulation is allowed by contact of nanomagnets with a noble metal layer and attributed here to the proximity effect phenomena at the interface. Another interesting feature observed is the decrease in the system disorder due to attenuation of nanomagnets interface defects. In this letter, we present experimental results obtained from direct $MFM$ measurements and support our findings numerically using the emergent model developed by some of us\cite{rodrigues2018towards}. 
We have performed our experimental observations in Permalloy square $ASI$ samples with same geometry and thickness used in previous work\cite{de2020effects}, with the addition of 20nm aluminum layer deposited by thermal evaporation on top of the $ASI$ geometry. Permalloy nanomagnets dimensions of $3\mu m\times400nm\times20nm$ are large enough to allow good magnetic force microscopy ($MFM$) signal and sufficiently small to preserve the range where single magnetic monodomains are observed. Sample Sq1 with lattice parameter a=3.95$\mu m$ and Sq2 with a=4.35$\mu m$ were chosen due to the presence of steps in hysteresis curve attributed to nanomagnets border defects that originates cracking reversal\cite{de2020effects}. 

\begin{figure}
    \centering
    \includegraphics[scale=0.25]{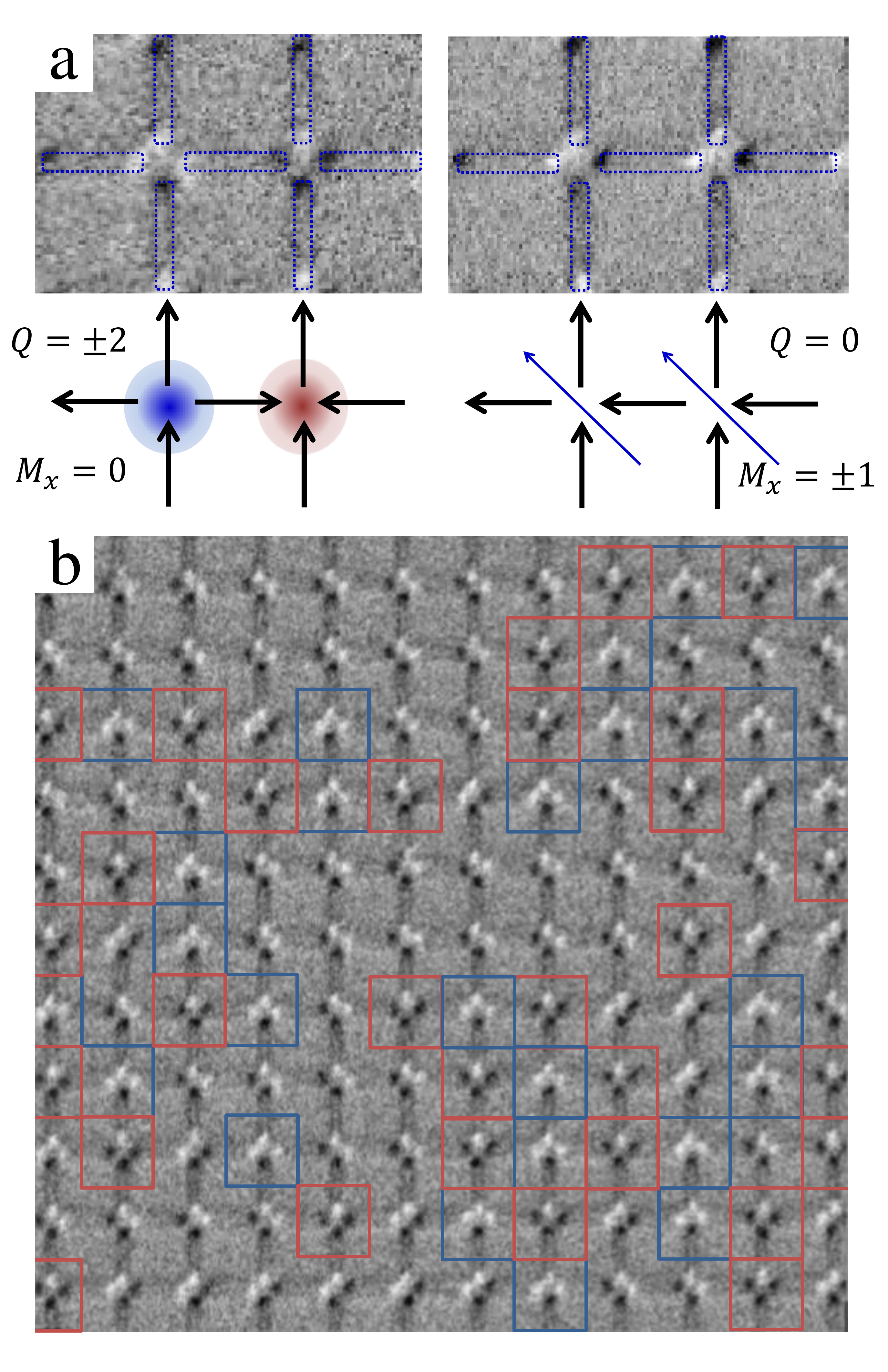}
    \caption{a) $MFM$ measurement at vertex showing respectively 3 out - 1 in / 3 in - 1 out configurations allowing emergence of charges $Q=+2$ and $Q=-2$, followed by 2 in - 2 out configuration of emergent dipole at vertex. b) Sample of measurement used to count magnetization and monopole density, realized during magnetization reversal process, with opposite charge monopoles highlighted by red and blue squares.}
    \label{fig:1}
\end{figure}

For the reversal process characterizations we have saturated nanomagnets magnetization in the x-axis direction and performed steps of external magnetic field to the saturation in the opposite direction. Using the $MFM$ images acquired after each field step, we were able to follow magnetization evolution and monopole density as a function of the external field. Ferromagnetic alignment of macro spins at a vertex give magnetization $M_{x}= \pm 1$ and zero charge, while antiferromagnetic alignment allow emergent charge $Q=\pm 2$ and null magnetization, as depicted in Figure 1a. In figure 1b we exemplify how experimental data was acquired. Measurements were performed in $50\mu m^2$ area comprising $12 \times 12$ vertices and the number of horizontal nanoislands pointing to left or right was counted as well as the number of emergent monopoles for each magnetic field step. In the figure, emergent monopoles are highlighted by red and blue squares according to its charge signal. The evolution of magnetization and monopole density as a function of the external magnetic field, performed in samples with and without aluminum interface, were systematically observed in several lattices investigated and are summarized by the results measured in Sq1 geometry presented in Figure 2. In the hysteresis curve of Figure 2a, it is possible to observe reasonable coercivity decrease and vanishing of steps related to magnetostatic traps from nanomagnets defects when aluminum layer is present. Similar results were reported using vibrating sample magnetometry technique for $ASI$ in contact with normal metal and superconductor\cite{kaur2017magnetic}.

\begin{figure*}[hbt!]
    \centering
    \includegraphics[scale=0.18]{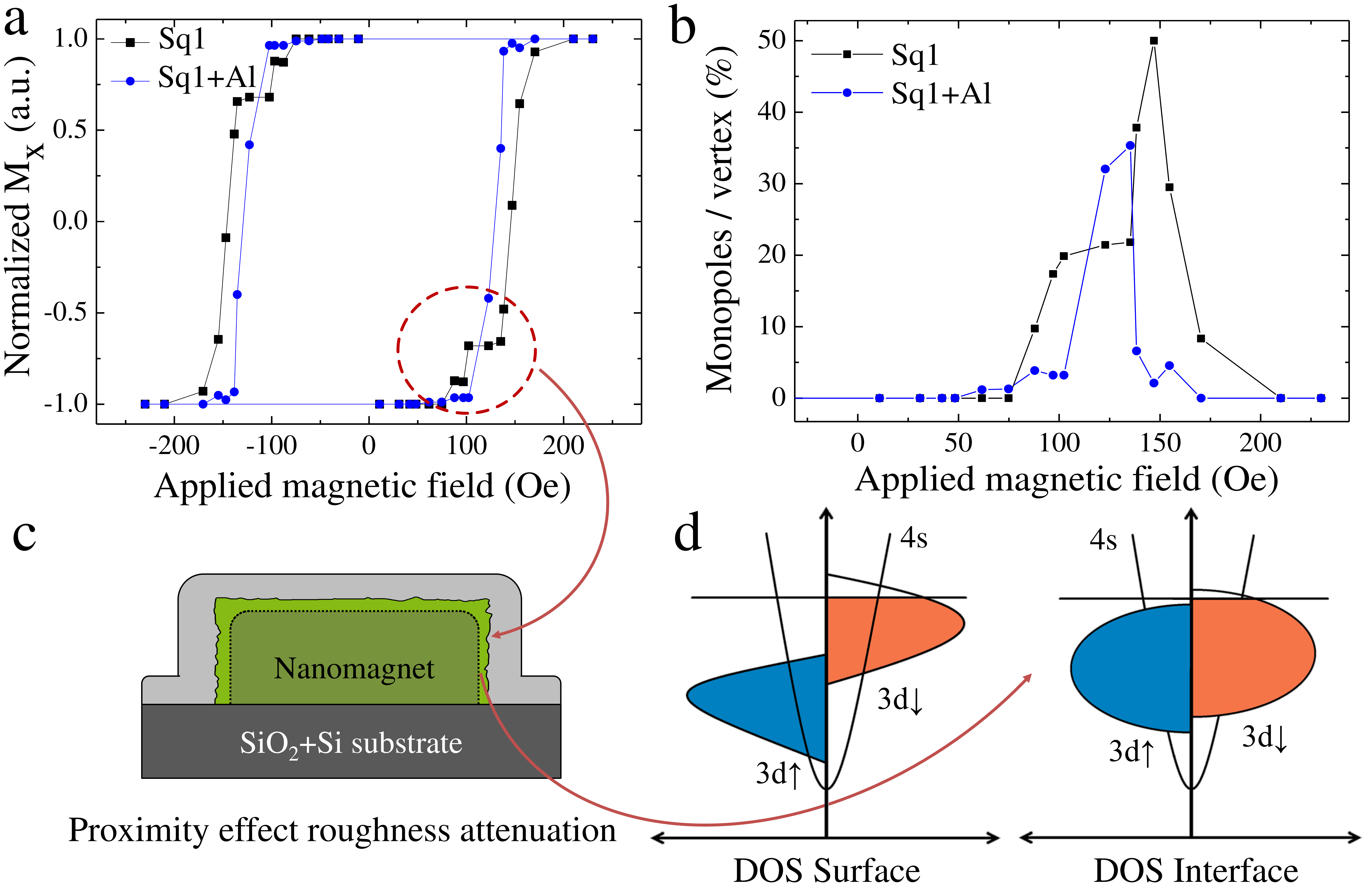}
    \caption{a) Hysteresis curve obtained by magnetization count at vertex during reversal process for Sq1 $ASI$ sample and $ASI$ samples covered by 20nm aluminum Sq1+Al, b) Density of monopole evolution during reversal process, c) suggested defect smoothness by proximity effect and d) scheme of modification in density of state (DOS) by $sp-d$ hybridization at the interface\cite{tersoff1982magnetic}. }
    \label{fig:2}
\end{figure*}

\begin{figure}
    \centering
    \includegraphics[width=0.5\textwidth]{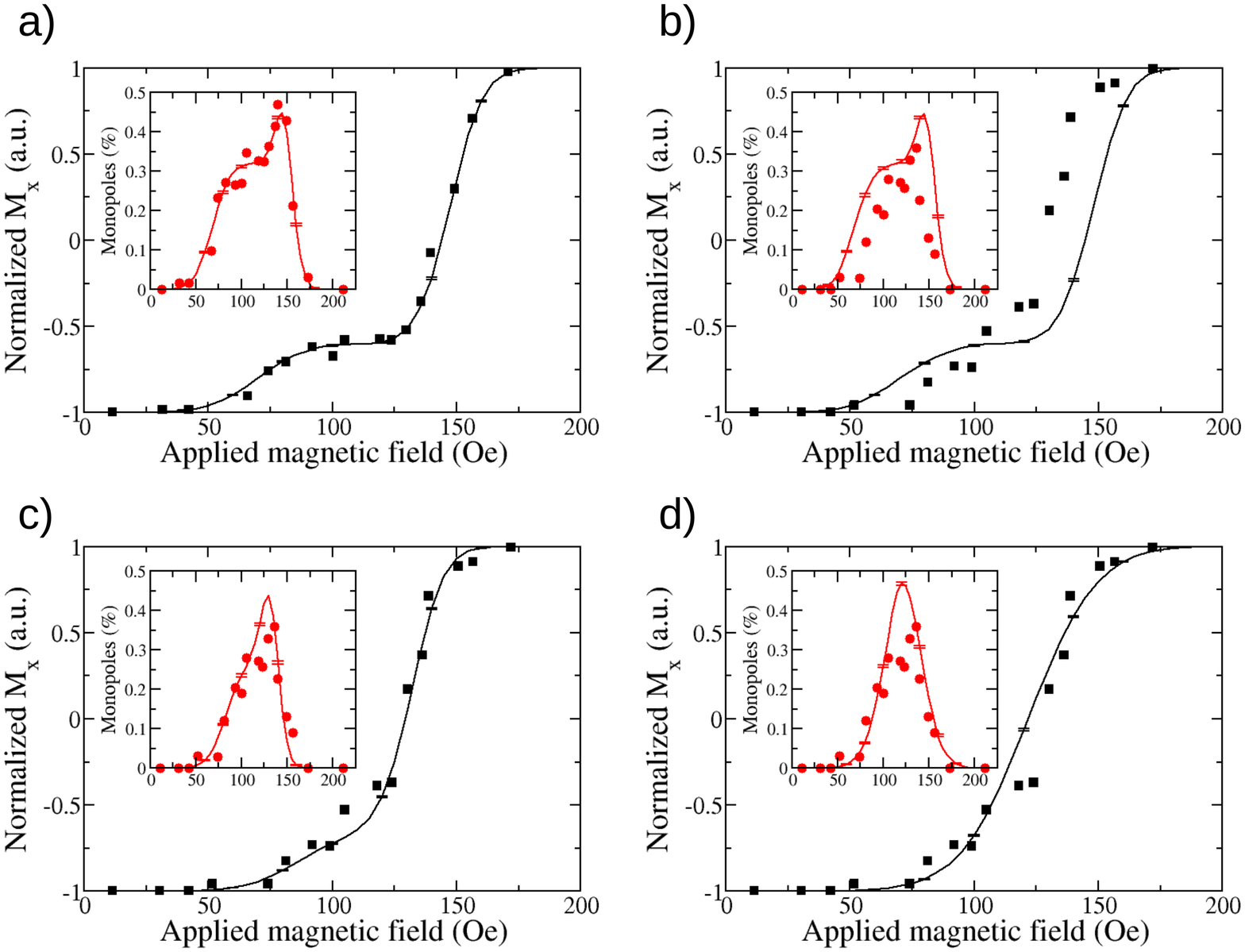}
    \caption{ Experimental data for sample Sq2 (black squares and red dots) for the hysteresis (outer plots) and magnetic monopole density (insets) compared to results for simulations (solid curves). In a) we show the curves for the sample without the aluminum cover. Simulational results were obtained using $q=0.5 \mu/a$, $M=2.8 \mu$, $h_c^{80}=160 Oe$, $\sigma^{80}=7.5\%$, $h_c^{20}=80 Oe$ and $\sigma^{20}=20\%$. In b), c) and d) the curve for the sample with the aluminum cover is compared to simulations using different distributions for the reversal field of the nanoislands. In b) we used $q=0.475 \mu/a$, $M=2.71 \mu$, $h_c^{80}=160 Oe$, $\sigma^{80}=7.5\%$, $h_c^{20}=80 Oe$ and $\sigma^{20}=20\%$. In c) $q=0.475 \mu/a$, $M=2.71 \mu$, $h_c^{82}=144 Oe$, $\sigma^{82}=8\%$, $h_c^{18}=96 Oe$ and $\sigma^{18}=10\%$. In d) $q=0.475 \mu/a$, $M=2.71 \mu$, $h_c^{100}=133 Oe$, $\sigma^{100}=18\%$.}
 
    \label{fig:3}
\end{figure}

Most pronounced effect can be noticed in the monopole density evolution curve of Figure 2b, with smaller percentile of monopoles being generated in a sharp shape curve for lowest magnetic field. Those phenomena promoted by aluminum interface are qualitatively similar to what was observed in samples with different lattice parameters \cite{de2020effects} where it was found that increasing distances among nanomagnets at the vertex reduces the stray field from the nanomagnets, affecting directly the magnetic monopole and dipole intensity. 
The present aluminum interface effect suggest fascinating interaction among metallic electron gas with emergent magnetic monopole and dipole, decreasing its intensity in a sort of electric screening. In order to elucidate the origins of such physical phenomena, we call on experimental observations of spin polarization decay in first atomic layers of ferromagnetic thin films in contact with noble metals in spin valve structures\cite{moodera1984magnetic}. Such magnetization suppression was theoretically investigated by first principle calculations and ascribed to hybridization of $sp-d$ orbitals at the interface\cite{tersoff1982magnetic,hong1991proximity}. Scattering of electrons from $sp$ to $d$ orbital modify the density of states shape, decreasing spin density unbalance at Fermi level of ferromagnetic materials, as depicted in the cartoon of Figure 2c, extremely decreasing spin polarization of first atomic layers. Such interface magnetization suppression would also explain the vanishing of magnetostatic traps related to nanomagnets defects, once it would be smoothed by the proximity effect as suggested in Figure 2d. Emergent magnetic monopole charge and dipole intensity are totally related to the dipolar interaction, that in turn is related to the nanoislands' dipole moment. The proximity effect with noble metals then would be responsible for intensity decrease of such emergent particles.      
To the best of our knowledge, experimental investigations or calculations of proximity effects by noble metals in nanostructures were not reported so far and in this letter we are going to use an emergent excitation model\cite{rodrigues2018towards} to investigate how the magnetization interface decrease in nanomagnets would imply in the observed behavior of magnetization and monopole density in the reversal processes of an ASI. In our previous work\cite{rodrigues2018towards,de2020effects}, numerical solution of a model of emergent excitations that considers the presence of magnetic monopoles of charge $q$ and magnetic dipoles of moment $|\vec{M}|$ on the vertices, presented a qualitative match with the evolution of magnetization and monopole density in reversal process of experimental square $ASI$ arrays\cite{de2020effects}. In this model\cite{rodrigues2018towards}, the total magnetic field at nanoisland $i$, $\vec{B}^{tot}_i$, is the sum of the field produced by the excitations on the vertices (see Refs.~\cite{rodrigues2018towards,de2020effects} for details). added to the external magnetic field. The nanoisland magnetization (considered as a monodomain pointing in direction $\vec{S}_i$) is flipped when $\vec{B}^{tot}_i\cdot\vec{S}_i < -h_i$, where $h_i$ is an intrinsic switching constant of nanoisland $i$. In order to take into account differences among nanoislands, the $h_i$ values are drawn from a Gaussian distribution centered at $h_c$ with standard deviation $\sigma_d=\sigma h_c$. In our previous work\cite{de2020effects}, we found that in order to qualitatively  describe experimental results, a kind of bimodal distribution had to be used, in such a way that for 90\% of the nanoislands, we used $h_c=h_c^{90}$ and $\sigma^{90}$ and for the remaining 10\% of the islands different values, $h_c=h_c^{10}$ and $\sigma^{10}$ were used. In general, $h_c^{x}$ will represent the mean switching field for $x\%$ of the nanoislands and $\sigma^{x}h_c^{x}$ the standard deviation of the gaussian distribution of switching fields for the same $x\%$ of the nanoislands.
In the figure \ref{fig:3} experimental data for sample Sq2 is compared to simulational results. First, in figure 3a, we compare results for the sample without the aluminum cover. The outer plot shows part of the hysteresis curve and the inset shows the evolution of the magnetic monopoles density. The symbols represent the experimental data while solid curves are results from simulations averaged over 50 different distributions of switching fields. For the charge and dipole strengths we used the results we obtained in our previous work~\cite{de2020effects}, $q=0.5 \mu/a$ and $M=2.8\mu$. The reversal field distribution was manually adjusted to fit the experimental curves. The best results were obtained for $h_c^{80}=160 Oe$, $\sigma^{80}=7.5\%$, $h_c^{20}=80 Oe$ and $\sigma^{20}=20\%$. As can be seen, the agreement is remarkable! 

The aluminum cover is expected to cause interface effects, leading to polarization zero around 0.7nm deep in nickel thin film \cite{moodera1984magnetic} (present in majority of Permalloy alloy used here). Thus, considering an effective thickness for the nanoislands of 19.3 nm instead of the original 20 nm, the effective volume would be reduced by approximately 5\%, in such a way that a reduction of the same amount would be expected for the charge, $q$, and dipole moment, $M$, of emergent excitations. Figure 3b shows the comparison among experimental results for the Sq2 sample with aluminum cover and simulational results considering $q=0.475 \mu/a$, $M=2.71 \mu$ and the same reversal fields used for the sample without the aluminum cover, $h_c^{80}=160 Oe$, $\sigma^{80}=7.5\%$, $h_c^{20}=80 Oe$ and $\sigma^{20}=20\%$. As can be seen, the results strongly suggest that the biggest effect of the aluminum cover will be on the switching fields and possibly on the reversal mechanism of the nanoislands. Interesting is the fact that the best adjust we achieved was obtained by changing the switching field parameters by about 10\%, as if they depend on the square of the volume. Figure 3c shows the comparison between experimental data and simulations performed using $q=0.475 \mu/a$, $M=2.71 \mu$, $h_c^{82}=144 Oe$, $\sigma^{82}=8\%$, $h_c^{18}=96 Oe$ and $\sigma^{18}=10\%$. Now the bimodal distribution of switching fields approaches a unimodal distribution since the mean switching fields approaches each other. The heterogeneity present in the sample was reduced! Indeed, to give further support to this assertive, reasonable results were also obtained by considering a single Gaussian (unimodal) distribution centered at $h_c^{100}=133 Oe$ with $\sigma^{100}=18\%$ as can be seen in figure 3d, reinforcing the softening effect of the aluminum layer.

In summary we have investigated by direct $MFM$ measurements, performed during a magnetization reversal process of $ASI$ samples, the evolution of emergent magnetic monopoles and dipoles in samples with and without the interface effect of a 20nm aluminum cover. Results show a systematic decrease in the coercivity and lowering of monopole creation density. The sharpening of the monopole density curve is related to a decrease in the heterogeneity of the distribution of the nanoislands switching fields and to a decrease on monopoles charge and dipoles moment. That decrease was attributed here to the annihilation of magnetostatic traps present on the nanomagnets surface by the magnetization vanishing in the first atomic layers of the nanomagnet interface and to the reduction of the overall magnetic moment of the nanoislands. Our simulation using an emergent model taking into account the value of nanomagnetic dipole under proximity effect supported our findings with very good accuracy. We believe that utilization of such covering may lead to more homogeneous systems, specially in what refers to the switching field of the nanoislands. In addition, in thinner samples it could be utilized for thermodynamic activation and the utilization of partial covering in different geometries could allow investigations of samples with different monopole interaction and mobility.      

\begin{acknowledgments}
We wish to acknowledge the discussions with Jagadeesh Moodera and would like to thank the Brazilian agencies CNPq, FAPEMIG
and Coordena\c{c}\~{a}o de Aperfei\c{c}oamento de Pessoal de
N\'{i}vel Superior (CAPES) - Finance Code 001.
\end{acknowledgments}

\bibliography{Al-ASI}

\end{document}